\documentstyle[aps,epsfig,prl]{revtex}
\begin{document}
\title{A Role of a Spatial Dispersion of an Electromagnetic Wave Coming Through a
Quantum Well }
\author{ L. I. Korovin, I. G. Lang}
\address{A. F. Ioffe Physical-Technical Institute, Russian
Academy of Sciences, 194021 St. Petersburg, Russia}
\author{D. A. Contreras-Solorio, S. T. Pavlov\cite{byline1}}
\address {Esculin de Fisica de la UAZ, Apartado Postal C-580,
98060 Zacatecas, Zac., Mexico}

\twocolumn[\hsize\textwidth\columnwidth\hsize\csname @twocolumnfalse\endcsname
\date{\today}
\maketitle \widetext
\begin{abstract}
\begin{center}
\parbox{6in} {A theory of light transmission through a quantum well (QW) in a magnetic
field perpendicular to the QW plane is developed. The light wave length is supposed
comparable with the QW width. The formulas for reflection, absorption and
transmission take into account the  spatial dispersion of the light monochromatic wave
and a difference of the refraction indexes of the QW and barrier. We suppose a normal
light incidence on the QW plane and consider only one excited energy level. These two
factors influence mostly light reflection, since an additional reflection from the QW
borders appears to the reflection due to interband transitions in the QW. The most
radical changes in reflection appear when a radiative broadening of the excited
energy level is small in comparison to a nonradiative broadening. Our theory is
limited by the condition of existence of size-quantized energy levels which is
satisfied for quite narrow QW's.}

\end{center}
\end{abstract}

] \narrowtext

Reflection and transmission data of an electromagnetic wave coming through a QW
contain some characteristic features carrying a valuable information about electronic
processes in  QW's \cite{1,2,3,4}. Most interesting results are available for
electronic systems with discrete energy levels. Such a situation is realized in a QW
in a quantizing magnetic field perpendicular to the QW plane or for excitonic energy
levels in absence of magnetic field. Modern semiconductor technologies allow to obtain
QW's of high quality where radiative broadenings of absorption lines may be
comparable with nonradiative broadenings or even exceed them. In such cases one cannot
use only the linear approximation on the electron-photon interaction, but must take
into account all the orders of this interaction
\cite{5,6,7,8,9,10,11,12,13,14,15,16,17,18}.

Reflection, absorption and transmission of an electromagnetic wave interacting with
QW discrete energy levels in an interband light frequency region had been considered
in \cite{12,13,14,15,16,17,18}. Both light pulses\cite{12,13,14,15,16,17} and
monochromatic radiation\cite{18} had been considered. One\cite{16}, two\cite{17,18}
and many excited energy levels\cite{15} had been taken into account. These results are
rightful for comparatively narrow QW's, when the inequality

\begin{equation}
\label{1} \kappa\, d<<1
\end{equation}
is fulfilled ( $d$ is the QW's width, and $\kappa$ is the modulus of the wave vector
${\bf k}$ of the light wave). The parameter $\kappa\, d$ was equalled to zero in
mentioned above articles, thus obtained reflection, absorption and reflection did not
depend on the QW's width  $d$.

Let us make use the wave length $0,8\,\mu$ (a laser on the base of GaAs) to estimate
numerically the value $\kappa$. The energy, corresponding to this wave length, is
$\hbar\, \omega_{\,l\,} =1.6\,eV$. If the refraction index of the QW material is $\nu
= 3.5$, then $\kappa=\nu\,\omega_{\,l\,}/c = 2.8 10^5\,cm^{-1}$ ($c$ is the light
velocity in vacuum). For the QW width $d=500 A$ the parameter $\kappa\,d=1.4$. Thus,
taking into account the spatial dispersion may be essential for the wide enough QW's.

For wide QW's the inequality $d\gg a_0$ (where $a_0$ is the lattice constant) is very
strong, thus one can apply the Maxwell equations for the continuum to describe a
transmission of light wave through a QW. Properly speaking, at such approach one has
to take into account a difference in refraction indexes of the QW and barrier. Then
some additional reflection must appear from  QW's borders which will decrease with
decreasing of the parameter $\kappa\, d$, but in the region $\kappa\, d\simeq 1$ it
may become equal or even enhance the reflection due to interband transitions inside of
the QW. Transmission of the light wave will be also changed. Thus the difference in
refraction indexes of the QW and barrier must be taken into account as well as the
spatial dispersion of the light wave.

Our article is devoted to a consideration of influence of both factors on reflection,
transmission and absorption of an electromagnetic wave coming through a QW and
exciting interband transitions in the QW.

\section{The model and main relations}

Let us consider a deep semiconductor QW displaced in the interval $0\leq z\leq d$
between two half-infinite barriers. A constant quantizing magnetic field is directed
perpendicularly to the QW's plane (along the z axis). An external light wave
penetrates along the $z$ axis from the negative $z$. We suppose that barriers are
transparent for the light wave, which is absorbed in part inside of the QW, inducing
interband transitions. At $T=0$ (the ground state) the valence band is filled
completely and the conductivity band is empty. In an excited state at least one
electron is transited into the conductivity band from the valence band where one hole
remained. This is true in a linear approximation on the wave amplitude. Let us
consider the light frequencies which are close to the QW band gap, when a small part
of the valence electrons near the band extremum takes part in absorption and the
effective mass approximation is applicable. For deep QW's one can neglect by electron
tunneling into barriers and consider the QW-barrier border as very sharp one (a
current is absent inside barriers). Energy levels near the QW bottom can be considered
in an approximation of an infinitely deep QW, though this approximation is not a
principal one and our theory may be extended for QW's of finite depths. Our system is
heterogeneous one. Since the heterogeneity size (the QW in our case) is smaller or of
the order of the light wave length optical characteristics of such system are to be
determined from a solution of Maxwell equations with current and charge densities
obtained with the help of the microscopic consideration\cite{19,20}.

The final result will be obtained for the only discrete energy level for the
electronic system in a QW. We can neglect an influence of other energy levels if the
exciting light frequency $\omega_{\,l\,}$ is close to the interband transition
frequency $\omega_{\,0\,}$. In the case $\hbar {\bf K}_\perp=0$ ($\hbar {\bf
K}_\perp$ is the electron-hole (EHP) summary quasi-momentum in the QW plane) discrete
energy levels in the QW are excitonic energy levels or the Landau levels in a
quantizing magnetic field directed perpendicularly to the QW plane. We consider an EHP
energy level in a quantizing magnetic field. The Coulomb interaction is a weak
perturbation for strong magnetic fields and not too wide QW's and it is neglected
below\cite{21}. However the excitonic effect will not lead to principal changes of
our results: it will change only the radiative broadening $\large {\gamma_{\,r\,}}$
of the electronic excitation. This is also true  for excitonic energy levels in
absence of magnetic fields.

Let us calculate the high-frequency current density in the QW induced by the exciting
light. For any spatially heterogeneous system one can introduce the conductivity
tensor $\sigma_{{\,\alpha\,\beta\,}}({\bf k},\omega|{\bf r})$ which links the average
current density $\,{\bf J}({\bf r},t)$ and the electric field

\begin{eqnarray}
\label{2}
J_{\,\alpha\,}({\bf r},t)=(2\pi)^{-4}\int d\,{\bf k}\int_0^\infty d\,\omega
 \sigma_{{\,\alpha\,\beta\,}}({\bf k},\omega|{\bf r})\,\nonumber\\
\times E_{\,\beta\,} ({\bf k},\omega)\exp [{i({\bf k\,r}-\omega \,t)}]+c.c.,
\end{eqnarray}
\begin{equation}
\label{3}
 E_{\,\beta\,} ({\bf k},\omega)=\int d\,{\bf r}\int_{-\infty}^\infty d\,t
 E_{\,\beta\,}({\bf r},t) \exp [{-i({\bf k\,r}-\omega\, t)}].
\end{equation}
Since the temperature $T=0$ the current density is averaged on the ground state of
the electronic system

\begin{equation}
\label{4}
 {\bf J}({\bf r},t)=<0|\,\hat {\bf J}({\bf r},t)\,|0>,
\end{equation}
where $\hat {\bf J}({\bf r},t)$ is the current density operator in linear
approximation on the external field.

We make use the following expression for the conductivity tensor  \footnote
{\normalsize {In Eq. (5) contributions containing factors of the order
 $\large {\kappa \,v/\omega}$ are omitted ( $v$
is the electron velocity).}}
\begin{eqnarray}
\label{5} \sigma_{\,\alpha\,\beta\,}({\bf k},\omega|{\bf r})=(i/\hbar)\int d\,{\bf r}^
{\,\prime\,}\int_{-\infty}^\infty d\,t^{\,\prime\,}\,\nonumber\\
\times\Theta(t^{\,\prime\,}) <0\,|\,[j_{\,\alpha\,}({\bf r},t),\tilde
d_{\,\beta\,}({\bf r}- {\bf r}^{\,\prime\,},t-t^{\,\prime)\,}\,]\,|\,0>,
\end{eqnarray}
where $\Theta (t)$ is the Haeviside function, and $j_{\,\alpha\,}({\bf r},t)$ is the
projection of the current density without external magnetic field, but taking into
account the constant quantizing magnetic field
\begin{equation}
\label{6}
{\bf H}=rot {\bf A}^{\,(0)\,}({\bf r}),
\end{equation}
where ${\bf A}^{(0)}({\bf r})$ is its vector potential. This operator is as follows

\begin{equation}
\label{7}
 j_{\,\alpha\,}({\bf r},t)=\exp (i{\cal H}\,t/\hbar)j_{\,\alpha\,}({\bf r})
 \exp (-i{\cal H}\,t/\hbar),
\end{equation}
\begin{equation}
\label{8}
 j_{\,\alpha\,}({\bf r})=(e/2)\sum_i[\,v_{\,i\,\alpha}\,\delta({\bf r}-
 {\bf r}_{\,i\,})+ \delta({\bf r}-{\bf r}_{\,i\,})\,v_{\,i\,\alpha}\,],
 \end{equation}
\begin{equation}
\label{9}
 v_{\,i\,\alpha}=-i(\hbar/m_{\,0\,})\,\partial/\,\partial\,r_{\,i\,\alpha}\,-
 \,(e/m_{\,0\,}\,c)A_{\,\alpha\,}^{{\,(0)\,}}({\bf r}_{\,i\,}),
\end{equation}
${\cal H}$ is the Hamiltonian of the electronic system in a quantizing magnetic field
(but without an external electromagnetic field),

\begin{equation}
\label{10}
 \tilde d_{\,\alpha\,}({\bf r},t)=\exp (\,i\,{\cal H}\,t/\hbar)\;
 \tilde d_{\,\alpha\,}({\bf r}\,)\;\exp (-\,i{\cal H}\,t/\hbar),
\end{equation}
\begin{equation}
\label{11}
 \tilde d_{\,\alpha\,}({\bf r})=e\sum_i\,(r_{\,i\,\alpha\,}-
 <0\,|\,r_{\,i\,\alpha\,}
 \,|\,0>)\, \delta({\bf r}-{\bf r}_{\,i\,}^{\,\prime\,}).
\end{equation}

Applying the effective mass approximation and taking into account the homogeneity of
the system in the QW plane  from Eqs. (2), (5) we obtain that
\begin{eqnarray}
\label{12}
\bar J_{\,\alpha\,}(z,t)=\frac{i}{4\pi^{\,2\,}}\frac{(e/m_{\,0\,})^{\,2\,}}
{\hbar\,\omega_{\,g\,}a_{\,H\,}^{\,2\,}}\int_{-\infty}^\infty d\,\omega
e^{-i\,\omega\,t}\sum_\chi\,\Phi_{\,\chi\,}(z)\nonumber\\
\times \left [\frac{p_{\,c\,v\,\alpha\,}^{\,j\,*}\, p_{\,c\,v\,\beta\,}^{\,j\,}}
{\omega-\omega_{\,\chi\,}+i\gamma_{\,\chi\,}/2}+
\frac{p_{\,c\,v\,\alpha\,}^{\,j\,}\,p_{\,c\,v\,\beta\,}^{\,j\,*}}
  {\omega+\omega_{\,\chi\,}+i\gamma_{\,\chi\,}/2}
\right ]\nonumber\\
\times\int_{-\infty}^\infty d\,z^{\,\prime\,} \Phi_{\,\chi\,}(z^{\,\prime\,})
E_{\,\beta\,}(z^{\,\prime\,},\omega)
\end{eqnarray}
where $\bar J_{\,\alpha\,}(z,t)$ is the current density averaged on the ground state
of the electronic system. An additional averaging on the elementary cell is performed
(which is available under condition  $d\gg a_{\,0\,}$. Following designations are used
in Eq. (12): $m_{\,0\,}$ is the bare electron mass,
 $\hbar \, \omega_{\,g\,}$ is the energy gap,
 $a_{\,H\,}=(c\,\hbar/|e|\,H)^{\,1/2\,}$ is the magnetic length, $\chi$
 is the set of the indexes

\begin{equation}
\label{13}
\large {\chi=(j,\lambda),\quad \lambda=(n=n_{\,c\,}=
n_{\,v\,},m_{\,c\,},m_{\,v\,})},
\end{equation}
 $j$ is the valence band number (since the valence band in cubic crystals
 (considered below) is degenerated),
$\large {n_{\,c\,}(n_{\,v\,})}$ is the Landau quantum number, $\large
{m_{\,c\,}(m_{\,v\,})}$ is the size quantization quantum number,
\begin{equation}
\label{14}
\Phi_{\,\lambda\,}(z)=(2/d)\sin(\,\pi\,m_{\,c\,}\,z/d)
\sin(\,\pi\,m_{\,v\,}\,z/d)
\end{equation}
is the product of the electron and hole wave functions depending on $z$,
\begin{equation}
\label{15}
E_{\,\beta\,}(z,\omega)=\int_{-\infty}^\infty d\,t\exp(\,i\,\omega\,t)
E_{\,\beta\,}\,(z,t),
\end{equation}
$\large {{\bf p}_{\,c\,v}^{\,j\,}}$ is the interband matrix element of the
quasi-momentum, corresponding to the transition between extremums of the valence and
conductivity bands,
\begin{equation}
\label{16}
\large {\hbar\omega_{\,\lambda\,}=\hbar \omega_{\,g\,}+
\varepsilon (m_{\,c\,}) +
\varepsilon (m_{\,v\,}) + \hbar\,\Omega_{\,\mu\,}(n+1/2)}
\end{equation}
is the electronic excitation energy with indexes
 $\lambda$, \,$\varepsilon
(m_{\,c\,})(\varepsilon (m_{\,v\,}))$ is the size quantized energy of electrons
(holes), $\Omega_{\,\mu\,}=|e|\,H/\mu\,c$ is the cyclotron frequency, $\mu=m_{\,e\,}\
m_{\,h\,}/(m_{\,e\,}+m_{\,h\,})$ , $m_{\,e\,}(m_{\,h\,})$ is the electro (hole)
effective mass, and $\gamma_{\,\lambda\,}$ is the nonradiative broadening of the
excited state with the quantum numbers  $\lambda$. Obtaining Eq. (12) we used the
relation
\begin{equation}
\label{17}
{\bf r}_{\,c\,v\,}=-(i/m_{\,0\,}\omega_{\,g\,}){\bf p}_{\,c\,v\,},
\end{equation}
where ${\bf r}_{\,c\,v\,}$ is the interband matrix element of the radius-vector ${\bf
r}$. Eq. (12) is applicable for the monochromatic exciting wave as well as for light
pulses.

Further we use the model  (see\cite{15,16,17,18}) where the vectors ${\bf
p}_{\,c\,v\,}^{\,j\,}$ for two degenerated bands are as follows
\begin{eqnarray}
\label{18}
 {\bf p}_{\,c\,v\,}^{\,I\,}= p_{\,c\,v\,}({\bf e}_{\,x\,}-
 i\,{\bf e}_{\,y\,})/\sqrt 2, \nonumber\\ {\bf p}_{\,c\,v\,}^{\,II\,}=
 p_{\,c\,v\,}({\bf e}_{\,x\,}+ i\,{\bf e}_{\,y\,})/\sqrt 2,
\end{eqnarray}
where ${\bf e}_{\,x\,},{\bf e}_{\,y\,}$ are the unite vectors along the axis $x$ and
$y$, and $p_{\,c\,v\,}$ is the real constant. This model corresponds to heavy holes in
crystals with the zinc blend structure if the axis $z$ is directed along the fourth
order axis\cite{22,23}. If the circular polarization vectors are applied

\begin{equation}
\label{19} {\bf e}_{\,l\,}=({\bf e}_{\,x\,}\pm i\,{\bf e}_{\,y\,})/\sqrt 2,
\end{equation}
the conservation of the polarization vector is performed
\begin{eqnarray}
\label{20} \sum_{j=I,II}\left[\frac{{\bf p}_{\,c\,v\,}^{\,j\,*\,}({\bf e}_{\,l\,} {\bf
p}_{\,c\,v\,}^{\,j\,} )}{\omega - \omega_{\,\lambda\,} +
i\gamma_{\,\lambda\,}/2}+\frac{{\bf p}_{\,c\,v\,}^{\,j\,}({\bf e}_{\,l\,} {\bf
p}_{\,c\,v\,}^{\,j\,*\,})}{\omega + \omega_{\,\lambda\,} +
i\gamma_{\,\lambda\,}/2}\right]\nonumber\\
={\bf e}_{\,l\,}p_{\,c\,v\,}^{\,2\,}\left[\frac{1} {\omega - \omega_{\,\lambda\,}
+i\gamma_{\,\lambda\,}/2}+ \frac{1}{\omega + \omega_{\,\lambda\,}
+i\gamma_{\,\lambda\,}/2}\right],
\end{eqnarray}
where ${\bf e}_{\,l\,}$ is any vector of Eq. (19), what makes easy following
calculations. Since for the model Eq. (18) the projection $p_{\,c\,v\,z}=0$, the
current $\bar {\bf J}(z,t)$ is transverse one, the induced charge density
  $\rho (z,t)=0$. Then one can choose the gage  $\varphi (z,t)=0$,
  where $\varphi (z,t)$ is the scalar potential and
 $${\bf E}(z,t)=-(1/c)(\partial {\bf A}(z,t)/\partial t,\qquad
 {\bf H}(z,t)=rot\,{\bf A}(z,t),$$
 where ${\bf A}(z,t)$ is the vector potential of the electromagnetic wave.
Applying the relation
\begin{equation}
\label{21}
E_{\,\alpha\,} (z,\omega)=(i\omega/c)A_{\,\alpha\,} (z,\omega),
\end{equation}
we will transit in Eq. (12) to the vector potential. The result may be conveniently
written

\begin{eqnarray}
\label{22} \bar J_{\,\alpha\,}\,
(z,t)=-\frac{e_{\,l\,\alpha}\gamma_{\,r\,}\nu}{8\pi^{\,2\,}} \sum_\lambda
\Phi_{\,\lambda\,} (z)\int_{-\infty}^\infty\omega\,d\,\omega e^{-\,i\,\omega\,t}
\nonumber\\
\times\left[\frac{1}{\omega - \omega_{\,\lambda\,}
+i\gamma_{\,\lambda\,}/2}+ \frac{1}{\omega + \omega_{\,\lambda\,}
+i\gamma_{\,\lambda\,}/2}\right]
\nonumber\\
\times \int_0^ddz^{\,\prime\,\,}A(z^{\,\prime\,},\omega)\Phi_{\,\lambda\,}
(z^{\,\prime\,})+c.c.,
\end{eqnarray}
where
\begin{equation}
\label{23}
\gamma_{\,r\,}=(2e^{\,2\,}/\hbar\,c\,\nu)(p_{\,c\,v\,}^{\,2\,}
/m_{\,0\,}\hbar\,\omega_{\,g\,})(|e|H/m_{\,0\,}\,c)
\end{equation}
is the radiative broadening of the EHP energy level in a magnetic field under the
condition $\kappa\,d=0$ \cite{13,18}, and the scalar $A(z,\omega)$ is introduced:

\begin{equation}
\label{24}
{\bf A}(z,\omega)={\bf e}_{\,l\,}A(z,\omega)+
{\bf e}_{\,l\,}^{\,*\,} A^{\,*\,} (z,-\omega).
\end{equation}

\section{The electric field of the electromagnetic wave}
We proceed below from the two assumptions. First, the plane wave is a monochromatic
one with the frequency $\omega_{\,l\,}$, i. e. in Eq. (24)
\begin{equation}
\label{25} A(z,\omega)=2\pi\,\delta(\omega-\omega_{\,l\,})\,A(z),
\end{equation}
and the vector ${\bf A}(z,t)$ takes the form
\begin{equation}
\label{26}
{\bf A}(z,t)={\bf e}_{\,l\,}\exp(-i\,\omega_{\,l\,}t)A(z)\,+\,c.c..
\end{equation}
Second, the only excited energy level in the QW is taken into account. We assume that
rest energy levels are disposed  far away from this level and their influence is
negligible. The equation for the scalar amplitude $A(z)$ of the vector potential is

\begin{eqnarray}
\label{27} d^{\,2\,}A/dz^{\,2\,}+\kappa_{\,1\,}^{\,2\,}\,A=0,\nonumber\\
\kappa_{\,1\,}=\nu_{\,1\,}\omega_{\,l\,}/c,\quad z\leq 0,\ z\geq d,
\end{eqnarray}
where $\nu_{\,1\,}$ is the barrier refraction index. In the QW region  $0\leq z\leq
d$ we have the equation
\begin{equation}
\label{28}
d^{\,2\,}A/dz^{\,2\,}+\kappa^{\,2\,}A=-(4\pi/c)\bar J(z),
\end{equation}
where the scalar amplitude of the current density  $\bar J(z)$ for the case of the
only energy level according Eqs. (22), (25) is as follows
\begin{eqnarray}
\label{29} \bar J(z)=-(\gamma_{\,r\,}\,\nu\,\omega_{\,l\,}/4\pi)\Phi(z)
\int_0^dd\,z^{\,\prime\,} A(z^{\,\prime\,})\Phi(z^{\,\prime\,})\nonumber\\
\times[(\omega_{\,l\,}-\omega_{\,0\,}+i\gamma/2)^{-1}+
(\omega_{\,l\,}+\omega_{\,0\,}+i\gamma/2)^{-1}]\,\nonumber\\+\,c.c. .
\end{eqnarray}
For the sake of simplicity we have introduced designations
\begin{equation}
\label{30}
\Phi_{\,\lambda\,}(z)=\Phi(z),\quad \omega_{\,\lambda\,}=\omega_{\,0\,},
\quad \gamma_{\,\lambda\,}=\gamma.
\end{equation}
In the vicinity of the resonance  $\omega_{\,l\,}\,=\,\omega_{\,0\,}$ the term
proportional to  $(\omega_{\,l\,}+\omega_{\,0\,}+i\gamma/2)^{-1}$ in Eq. (29) is
omitted. Eq. (29) is integro-differential one. If one represents the solution of Eq.
(29) as a sum of the general solution of the homogeneous equation and the partial
solution of the non-homogeneous equation, then one obtains the Fredgholm integral
equation of the second type
 \footnote {\normalsize {The similar equation was considered in \cite{24}
 for the inversion layer.}}
\begin{eqnarray}
\label{31} A(z)=C_{\,1\,}e^{i\,\kappa\,z}+C_{\,2\,}e^{-i\,\kappa\,z}-
\frac{i\,(\,\gamma_{\,r\,}/2)F(z)}
{\omega_{\,l\,}-\omega_{\,0\,}+i\gamma/2}\nonumber\\
\times\int_{\,0\,}^{\,d\,} d\,
z^{\,\prime\,}\,A(z^{\,\prime\,})\,\Phi(z^{\,\prime\,}).
\end{eqnarray}
$C_1$ and $C_2$ are the arbitrary constants determined from boundary conditions in
the planes $z=0$ and $z=d$, and the function $F(z)$ is determined as
\begin{eqnarray}
\label{32} F(z)=\exp(i\,\kappa\,z)\int_{\,0\,}^{\,z\,}d\,z^{\,\prime\,}
\exp(-i\,\kappa\,z^{\,\prime\,})\,\Phi(z^{\,\prime\,})\nonumber\\+
 \exp(-i\,\kappa\,z)\int_{\,z\,}^{\,d\,}d\,z^{\,\prime\,}
\exp(i\,\kappa\,z^{\,\prime\,})\,\Phi(z^{\,\prime\,}).
\end{eqnarray}
If $\gamma_{\,r\,}\ll\gamma$, the integral term in Eq. (31) is a small perturbation
which may be counted in the first approximation only. If $\gamma_{\,r\,}\geq\gamma$,
one has to take into account the whole iterative sequence. Representing the desirable
function  $A(z)$ as a series
\begin{eqnarray}
\label{33} A(z)=A_{\,0\,}(z)+A_{\,1\,}(z)+A_{\,2\,}(z)+...,\nonumber\\
A_{\,0\,}(z)=C_{\,1\,}\exp(\,i\,\kappa\,z)+C_{\,2\,}\exp(-\,i\,\kappa\,z)
\end{eqnarray}
and substituting it in Eq. (31), we obtain the recurrent relation
$$A_{\,j\,}(z)=sF(z)\int_0^dd\,z^{\,\prime\,}\Phi(z^{\,\prime\,})
A_{\,j-1\,}(z^{\,\prime\,}),\quad j=1,2,3...$$ Making use this relation, one can
reduce Eq. (33) to the geometric progression
\begin{eqnarray}
\label{34} A(z)=A_{\,0\,}(z)-h\,s\,F(z)(1-s\,\varepsilon+
s^{\,2\,}\varepsilon^{\,2\,}-...)\nonumber\\=
A_{\,0\,}(z)-h\,s\,F(z)/(1+s\,\varepsilon),
\end{eqnarray}
where for the sake of simplicity we have designated
$$\int_0^dd\,z\,\Phi(z)A_{\,0\,}=h,$$
$$i(\gamma_{\,r\,}/2)/(\omega_{\,l\,}-\omega_{\,0\,}+i\gamma_{\,r\,}/2)=s$$ and have
introduced the complex function
\begin{equation}
\label{35}
\varepsilon = \varepsilon^{\,\prime\,}+i\varepsilon^{\,\prime\,\prime}=
\int_0^dd\,z^{\,\prime\,}\,\Phi(z^{\,\prime})\,F(z^{\,\prime\,}).
\end{equation}
Substituting  $h,\,s$ and $\varepsilon$ in Eq. (34) we obtain the solution
\begin{eqnarray}
\label{36} A(z)=C_{\,1\,}e^{i\,\kappa\,z}+C_{\,2\,}e^{-i\,\kappa\,z}\nonumber\\-
\frac{i\,(\gamma_{\,r\,}/2)\,F(z)}
{\omega_{\,l\,}-\omega_{\,0\,}+i(\gamma+\gamma_{\,r\,}\,\varepsilon)/2}\nonumber\\
\times\int_{\,0\,}^{\,d\,}d\,z^{\,\prime\,}\,(C_{\,1\,}\,
e^{i\,\kappa\,z^{\,\prime\,}}+C_{\,2\,}\,e^{-i\,\kappa\,z^{\,\prime\,}})
\,\Phi(z^{\,\prime\,}).
\end{eqnarray}
The complex value $\varepsilon$ determines broadening and the level shift which
appear due to the spatial dispersion of the light wave. As it follows from the
definition Eq.
 (35), in the limiting case  $\kappa\,d=0
 \;~~\varepsilon = \delta_{\,m_{\,c\,}\,m_{\,v\,}\,}$, and the integral in the RHS
 of Eq. (36) equals to $(C_{\,1\,}+C_{\,2\,})
 \delta_{\,m_{\,c\,}\,m_{\,v\,}\,}$, i. e.
 only permitted transitions $m_{\,c\,}=m_{\,v\,}$ contribute into the current.
 If  $\kappa\,d\neq 0$, the forbidden transition $m_{\,c\,}\neq m_{\,v\,}$
provides the interband current and the appearance in the denominator of Eq.
 (36) the value
  $\varepsilon$, however this $\varepsilon \to 0$ if
 $\kappa\,d\to 0$ . Let us also note  that the function Eq. (32)
 $F(z)=\delta_{\,m_{\,c\,}\,m_{\,v\,}\,}$ if $\kappa\,d\to 0$.
 Below we consider only the permitted transitions.

 The  solution of Eq. (27) is
\begin{eqnarray}
\label{37} A^{\,l\,}(z)&=&A_{\,0\,}\exp(i\,\kappa_{\,1\,}\,z )+C_{\,R\,}\,
\exp(-i\,\kappa_{\,1\,}\,z),\quad z\leq 0,\nonumber\\
A^{\,r\,}(z)&=&C_{\,T\,}\exp(i\,\kappa_{\,1\,}\,z),\quad z\geq d,
\end{eqnarray}
$C_{\,R\,}$, $C_{\,T\,}$ determine the amplitude of the reflected and transmitted
wave, respectively. On boundaries  $z=0$ and $z=d$ the continuity of the magnetic
field of the wave leads to the continuity of $dA/dz$, the continuity of the tangential
projections of the electric field leads to the continuity of $A(z)$. As a result we
obtain following expressions for the coefficients $C_{\,1\,},C_{\,2\,},C_{\,R\,}$ and
$C_{\,T\,}$:
\begin{eqnarray}
\label{38}
 C_{\,i\,}=A_{\,0\,}{\cal C}_{\,i\,}\quad(i=1,2), \quad
C_{\,R\,(\,T\,)}=A_{\,0\,}{\cal C}_{\,R\,(\,T\,)},\nonumber\\
{\cal C}_{\,1\,}=(2/\Delta)\exp(-i\,\kappa\,d)\,[1+\zeta+ (1-\zeta){\cal
N}\,],\nonumber\\
{\cal C}_{\,2\,}=-(2/\Delta)(1-\zeta)\, [\exp(i\,\kappa\,d)+{\cal N}],\nonumber\\
{\cal C}_{\,R\,}=\rho/\Delta,\nonumber\\
 {\cal C}_{\,T\,}=
4\,\zeta\exp{(-i\,\kappa_{\,1\,}\,d)}\, [1+\exp{(-i\,\kappa\,d)}\,{\cal N}]/\Delta,
\end{eqnarray}
\begin{eqnarray}
\label{39}
 \Delta=(\zeta+1)^{\,2\,}\exp(-i\,\kappa\,d)\nonumber\\-
(\zeta-1)^{\,2\,}\exp(i\,\kappa\,d)\nonumber\\
-2\,(\zeta-1)\,{\cal N}\,
[(\zeta+1)\exp(-i\,\kappa\,d)+\zeta-1]\nonumber\\
\rho=2i\,(\zeta^2-1)\sin\kappa\,d\nonumber\\
+2\,[(\zeta^{\,2\,}+1)\exp{(-i\,\kappa\,d)}+
\zeta^{\,2\,}-1]\,{\cal N}
\end{eqnarray}
In Eqs. (38), (39) we have introduced the determinations
\begin{equation}
\label{40} \zeta=\kappa\,/\,\kappa_{\,1\,}=\nu\,/\,\nu_{\,1\,},
\end{equation}
\begin{eqnarray}
\label{41} {\cal
N}=-s\,F^{\,2\,}(0)\nonumber\\=-{i(\gamma_{\,r\,}/2)\,F^{\,2\,}(0)/over
[\omega_{\,l\,}-\omega_{\,0\,}+i(\gamma+\gamma_{\,r\,}\varepsilon)/2]}.
\end{eqnarray}
Returning to the time-representation and transiting in Eq. (37) from $A(z)$ to the
electric fields to the left $(E^{\,l\,}(z,t))$ and to the right $(E^{\,r\,}(z,t))$ of
the QW we obtain that
\begin{equation}
\label{42}
{\bf E}^{\,l\,}(z,t)={\bf e}_{\,l\,}E_{\,0\,}e^{-i\,\omega_{\,l\,}\,t}\left
[e^{i\,\kappa_{\,1\,}\,z}
+{\cal C}_{\,R\,}e^{-i\,\kappa_{\,1\,}\,z}\right]+c.c.\ ,
\end{equation}
\begin{equation}
\label{43}
{\bf E}^{\,r\,}(z,t)={\bf e}_{\,l\,}E_{\,0\,}{\cal C}_{\,T\,}
e^{-i\,(\,\omega_{\,l\,}t-\kappa_{\,1\,}\,z)}+c.c.\,.
\end{equation}
The electric field inside of the QW is determined by Eq. (36), if to substitute $A(z)$
 by $E(z)$ and to transit to the time-representation. Expressions for fields
 contain $\varepsilon$ and $F(0)$, and the field inside of the QW
 contains additionally  $F(z)$, which contributes into the space dependence of the field.
For the case $m_{\,c\,}=m_{\,v\,}=m\quad F(z)$ and $\varepsilon$ are as follows
\begin{eqnarray}
\label{44} F(z)=iB\{2-\exp(i\kappa\,z)-\exp{[i\kappa\,(d-z)}]\nonumber\\
-(\kappa\,d/\pi m)^{\,2\,}\sin^{\,2\,}(\pi\,m\,z/d)\},
\end{eqnarray}
\begin{eqnarray}
\label{45} F(0)=F(d)=iB[1-\exp(i\,\kappa\,d)],\nonumber\\
B=\frac{4\pi^{\,2\,}m^{\,2\,}}{\kappa\,d\,[4\pi^{\,2\,}m^{\,2\,}-
(\kappa\,d)^{\,2\,}]},
\end{eqnarray}
\begin{eqnarray}
\label{46} \varepsilon^{\,\prime\,}=F^{\,2\,}(0)\exp(-i\kappa\,d)=
4B^{\,2\,}\sin^{\,2\,} (\kappa\,d/2), \nonumber\\
\varepsilon^{\,\prime\,\prime\,}=2B[1-B\sin\kappa\,d-
3(\kappa\,d)^{\,2\,}/(8\pi^{\,2\,}m^{\,2\,}].
\end{eqnarray}
The functions  $\varepsilon^{\,\prime\,}$ and $\varepsilon^{\,\prime\, \prime\,}$
from the parameter $\kappa\,d$  are represented in Fig.1.

In the limiting case of an homogeneous medium
 $(\kappa_{\,1\,}=\kappa)$ we obtain
\begin{eqnarray}
\label{47} {\bf E}^{\,l\,}(z,t)={\bf e}_{\,l\,}\,E_{\,0\,}e^{-i\,\omega_{\,l\,}\,t}
\nonumber\\
\times\left[e^{i\,\kappa\,z}-\frac{i(\,\tilde \gamma_{\,r\,}/2)}
{\Omega+i\,\Gamma\,/2}\,e^{i\,\kappa\,(d-z)}\right]\nonumber\\ + c.c.\,,
\end{eqnarray}
\begin{eqnarray}
\label{48}
{\bf E}^{\,r\,}(z,t)={\bf e}_{\,l\,}\,E_{\,0\,}e^{-i(\,\omega_{\,l\,}\,t-\kappa\,z)}
\left[1- \frac{i(\,\tilde \gamma_{\,r\,}/2)}
{\Omega+i\,\Gamma\,/2}
 \right]\nonumber\\ + c.c.\,,
\end{eqnarray}
where the new designations are introduced:
\begin{eqnarray}
\label{49} \Omega=\omega_{\,l\,}-\omega_{\,0\,}-
\varepsilon^{\,\prime\,\prime\,}\gamma_{\,r\,}/2,\quad
\Gamma=\gamma+\tilde\gamma_{\,r\,},\nonumber\\
\tilde\gamma_{\,r\,}=\gamma_{\,r\,}\left |\int_0^dd\,z\exp{(\,i\,\kappa\,z)}
\Phi(z)\right |^{\,2\,}=\gamma_{\,r\,}\,\varepsilon^{\,\prime\,}.
\end{eqnarray}
 The field inside of the QW is
\begin{eqnarray}
\label{50} {\bf E}(z,t)={\bf e}_{\,l\,}\,E_{\,0\,}e^{-i\omega_{\,l\,}\,t} \left
[e^{i\kappa\,z}-\frac{i(\gamma_{\,r\,}/2)\,F(0)\,F(z)} {\Omega+i\,\Gamma \,/2}\right
]\nonumber\\ + c.c.\;.
\end{eqnarray}
The value $\gamma_{\,r\,}$ coincides with EHP radiative broadenings in a quantizing
magnetic field calculated in \cite{13,18} at  ${\bf K}_{\,\perp\,}=0$ and for the
arbitrary value $\kappa\,d$. Comparing Eqs. (47), (48) to corresponding expressions
in \cite{18} for fields to the left and right of the QW we find that in the case
$\kappa\,d\neq 0$ the value $\gamma _{\,r\,}$ is substituted by $\tilde
\gamma_{\,r\,}$, the shift on the value $\gamma _{\,r\,}\,\varepsilon^{\,\prime
\,\prime\,}$ appears and the additional factor $\exp{(i\,\kappa\,d)}$ appears in the
expression for the induced wave to the left from the QW. One can make sure that the
induced field to the left of the QW coincides with the induced field to the right of
the QW  if the value $d-z$ is substituted by $z$. It is seen from Eq. (46) and in
Fig. 1 that $\tilde \gamma_{\,r\,}$ decreases with growing  $\kappa\,d$. At
$\kappa\,d\gg 1\; ~~\tilde \gamma_{\,r\,}\to 0$, what corresponds to the transition
from the QW to the bulk crystal. In such a case the contribution of the only energy
level into induced fields and, consequently, into absorption and reflection approaches
to zero.

In the limiting case $\gamma_{\,r\,}=0$  from Eqs. (42),(43) one obtains the well
known solution for a monochromatic wave spreading in a medium containing a transparent
layer of another matter \cite{26}.

\section{Reflection, absorption and transmission
of an electromagnetic wave}

Thus according Eq. (42) the electric field vector of the reflected wave $\Delta {\bf
E}_{\,\l\,}^{\,l\,}\,(z,t)$ and of the circular polarization is as follows

\begin{equation}
\label{51} \Delta {\bf E}^{\,l\,}\,(z,t)= {\bf e}_{\,\l\,} E_{\,0\,} {\cal
C}_{\,R\,}\,e^{-i\,(\,\omega_{\,l\,}\,t+ \kappa_{\,1\,}\,z)}+c.c. .
\end{equation}
According Eq. (43) the electric field vector of the transmitted  wave is

\begin{equation}
\label{52}
{\bf E}^{\,r\,}\,(z,t)={\bf e}_{\,\l\,}E_{\,0\,} {\cal C}_{\,T\,}
\,e^{-i\,(\,\omega_{\,l\,}\,t-\kappa_{\,1\,}\,z)}+c.c.\;.
\end{equation}

Analogically to \cite{18} let us introduce the part of the reflected energy $\cal R$
which is defined as the relation of the reflected flux energy module to the incident
energy flux module, i. e.

\begin{equation}
\label{53}
{\cal R}=|{\cal C}_{\,R\,}|^{\,2\,}.
\end{equation}
The part of the transmitted energy  $\cal T$ equals
\begin{equation}
\label{54}
{\cal T}=|{\cal C}_{\,T\,}|^{\,2\,},
\end{equation}
and the part of the absorbed energy $\cal A$ is defined as
\begin{equation}
\label{55}
{\cal A}=1-{\cal R}-{\cal T}.
\end{equation}

At first let us consider an influence of the spatial dispersion on a frequency
dependence of reflection of a homogeneous medium. From Eqs. (38), (39) and (53)-(55)
we obtain
\begin{eqnarray}
\label{56} {\cal R}=\frac{(\tilde \gamma_{\,r\,}/2)^{\,2\,}}{\Omega^{\,2\,}+
\Gamma^{\,2\,}/4},\quad {\cal A}=\frac{\gamma\tilde\gamma_{\,r\,}/2}{\Omega^{\,2\,}+
\Gamma^{\,2\,}/4},\nonumber\\
 {\cal
T}=\frac{\Omega^{\,2\,}+\gamma^{\,2\,}/4}{\Omega^{\,2\,}+ \Gamma^{\,2\,}/4}.
\end{eqnarray}
These expressions coincide with those without the spatial dispersion. The difference
is in the substitution of the function $\tilde\gamma_{\,r\,}$
($\tilde\gamma_{\,r\,}\to\gamma_{\,r\,}$ при $\kappa\,d\to 0$) instead of the constant
$\gamma_{\,r\,}$ and the appearance of the function $\varepsilon^{\,\prime\,\prime\,}$
which determines the shift of the extremum of the corresponding curve and disappears
in the limit $\kappa\,d=0$.

The spatial dispersion is demonstrated more stronger in reflection in the case
$\gamma\gg\gamma_{\,r\,}$. Indeed, if $\gamma\ll\gamma_{\,r\,}$ the maximum of the
reflection curve from Eq. (56) ${\cal R}_{\,max\,}\cong 1$ and does not practically
depend of $\kappa\,d$. If $\gamma\gg\gamma_{\,r\,}$, then $\tilde\gamma_{\,r\,}$ in
the denominator of Eq. (56) gives a small contribution in the dependence from
$\kappa\,d$ and this dependence is determined by the function $\tilde\gamma_{\,r\,}$
in the numerator. But however ${\cal R}_{\,max\,}=
(\tilde\gamma_{\,r\,}/\gamma)^{\,2\,}\ll1$ . Just the contrary picture is realized
for transmission: at $\gamma\ll\gamma_{\,r\,}$ of the transmission curve ${\cal
T}_{\,min\,}=(\gamma/\tilde\gamma_{\,r\,})^{\,2\,}\ll 1$ and depends noticeably from
$\kappa d$ growing with increasing $\kappa\,d$. When $\gamma\gg\gamma_{\,r\,}$ then
${\cal T}\cong 1$ and is weakly dependent on $\kappa\,d$. The maximum of the
absorption peak in these limiting cases is equal ${\cal
A}_{\,max\,}=2\gamma/\tilde\gamma_{\,r\,}\ (\gamma\ll\gamma_{\,r\,})$ and ${\cal
A}_{\,max\,}= 2\tilde\gamma_{\,r\,}/\gamma\,(\gamma\gg\gamma_{\,r\,})$, in the both
limiting cases ${\cal A}_{\,max\,}\ll 1$, но ${\cal A}_{\,max\,}\gg{\cal
R}_{\,max\,}\ (\gamma\gg\gamma_{\,r\,})$ и ${\cal A}_{\,max\,}\gg{\cal T}_{\,min\,}\
(\gamma\ll\gamma_{\,r\,})$. The frequency dependence ${\cal R}$, ${\cal A}$, и ${\cal
T}$ for the limiting cases
 $\gamma\gg\gamma_{\,r\,}$ и $\gamma\ll\gamma_{\,r\,}$ is represented in Figs.
 2-4. It is seen in Fig. 2 how the spatial dispersion influences the height and width
 of the reflection peak which decrease with growing $\kappa\,d$. The peak frequency shift
 is  inobservable  since $\varepsilon^{\prime\,\prime}\gamma_r\ll\gamma$.
Vice verse, in Fig. 3 the spatial dispersion leads to the shift of the reflection peak
without changing its form. Fig. 3 demonstrates also transmission ${\cal T}$ for the
case $\gamma\ll\gamma_{\,r\,}$. There are both the shift ${\cal T}_{\,min\,}$
 with growing $\kappa\,d$ and its increasing which is almost inobservable due to the
 chosen scale on the ordinate axis.
In Fig. 4 absorption ${\cal A}$ is demonstrated for two limiting cases. On the narrow
peak set  (Fig. 4b, the case
 $\gamma\ll\gamma_{\,r\,}$) one can see  growing of  ${\cal A}_{\,max\,}$, as well as
 its shift. The shift appearance (as well as in Fig. 2 for ${\cal T}$)
is due to that
  ${\cal A}_{\,max\,}\sim \tilde\gamma_{\,r\,}^{\,-\,1\,}$ and
 $\varepsilon^{\prime\,\prime}\gamma_{\,r\,}\cong\tilde\gamma_{\,r\,}$.
 The curves in Fig. 4а correspond to the case  $\gamma\gg\gamma_{\,r\,}$, the shift
 is small and
${\cal A}_{\,max\,}$ decreases with growing $\kappa\,d$.

If one takes into account a heterogeneity of the medium, i. e. $\zeta\neq
1\;(\nu\neq\nu_{\,1\,})$, and neglects the spatial dispersion then instead of Eq.
(56)one obtains the expressions

 \begin{eqnarray}
 \label{57}
 {\cal R}=\frac{\zeta^{\,2\,}(\gamma_{\,r\,}/2)^{\,2\,}}{\Omega^{\,2\,}+
 (\gamma+\zeta\,\gamma_{\,r\,})^{\,2\,}/4},\nonumber\\
 {\cal A}=\frac{\zeta\,\gamma\,\gamma_{\,r\,}/2}{\Omega^{\,2\,}+
 (\gamma+\zeta\,\gamma_{\,r\,})^{\,2\,}/4},\nonumber\\
 {\cal T}=\frac{\Omega^{\,2\,}+\gamma^{\,2\,}/4}{\Omega^{\,2\,}+
 (\gamma+\zeta\,\gamma_{\,r\,})^{\,2\,}/4}.
 \end{eqnarray}
 It is seen from Eq. (57) that in the limiting case
 instead of  $\gamma_{\,r\,}$ the value
 $\zeta \,\gamma_{\,r\,}$ (see Eq. (23)) figures
 in which the refractive index $\nu_{\,1\,}$ refers to the barrier matter.
 This coincide with the result of \cite{18}.

 \section{The general case}

 In this section the general situation is considered
 when the matter is heterogeneous
 and the spatial dispersion is essential.
 Making use Eqs. (38), (39) and (41)
 we reduce reflection to the expression

 \begin{equation}
 \label{58}
{\cal R}={\frac{(\tilde\gamma_{\,r\,}/2)^{\,2\,}\,X_{\,1\,}}
{\Omega^{\,2\,}+\Gamma^{\,2\,}/4}+v_{\,1\,}-
\frac{(\tilde\gamma_{\,r\,}/2)\,(Y_{\,1\,}\,\Omega+Z_{\,1\,}\,\Gamma/2)}
{\Omega^{\,2\,}+\Gamma^{\,2\,}/4}\over|\Delta|^{\,2\,}},
 \end{equation}
 \begin{equation}
 \label{59}
 |\Delta|^{\,2\,}=v+\frac{(\tilde\gamma_{\,r\,}/2)^{\,2\,}\,X-
 (\tilde\gamma_{\,r\,}/2)\,(\,Y\,\Omega+Z\,\Gamma/2)}
 {\Omega^{\,2\,}+\Gamma^{\,2\,}/4},
 \end{equation}
 where
 \begin{eqnarray}
 \label{60}
 v=4\,\zeta^{\,2\,}\cos^{\,2\,}\kappa\,d+
 (\zeta^{\,2\,}+1)^{\,2\,}\sin^{\,2\,}\kappa\,d,\nonumber\\
 v_{\,1\,}=(\zeta^{\,2\,}-1)^{\,2\,}\sin^{\,2\,}\kappa\,d,
 \end{eqnarray}
 \begin{eqnarray}
 \label{61}
 X=2\,(\,\zeta-1)^{\,2\,}[\,\zeta^{\,2\,}+1+
 (\zeta^{\,2\,}-1)\cos\kappa\,d\,],\nonumber\\
 X_{\,1\,}=2\,[\,\zeta^{\,4\,}+1+(\zeta^{\,4\,}-1)\cos\kappa\,d\,],
 \end{eqnarray}
 \begin{eqnarray}
 \label{62}
 Y=2\,(\,\zeta-1)^{\,2\,}(\zeta+1)\,\nonumber\\
 \times[\,(\,\zeta+1)\cos\kappa\,d+\zeta-1\,]\,
 \sin\kappa\,d,\nonumber\\
 Y_1=2\,\zeta^{\,2\,}\,(\,\zeta^{\,2\,}-1)\,\sin\kappa\,d,
 \end{eqnarray}
\begin{eqnarray}
 \label{63}
 Z=2\,(\,\zeta-1)\,(\,\zeta-1)\,(\,\zeta+1)^{\,2\,}
 \sin^{\,2\,}\kappa\,d\nonumber\\-
 2\,\zeta\,[\,(\,\zeta+1)\cos\kappa\,d+\zeta-1\,],\nonumber\\
Z_{\,1\,}=2\,(\,\zeta-1)^{\,3\,}(\zeta+1)\sin\kappa\,d.
 \end{eqnarray}
 In the denominator of the function ${\cal R}$, determined by Eq. (59),
 the function $v\,(v\gg 1)$ gives the main contribution, the rest terms may be neglected
 . One exclusion is the term
 $\sim\Gamma$, which contributes essentially in the case
 $\gamma\ll\tilde\gamma_{\,r\,}$. In the numerator  ${\cal R}$ the function
 $v_{\,1\,}$ determines reflection from the QW borders. This reflection does not
 depend from the light frequency in the frequency interval corresponding to the peak
 width and disappears, as it is seen from Eq. (60), in the limiting cases $\kappa\,d\to 0$
 and $\zeta\to 0$.
 At $\gamma\ll\tilde\gamma_{\,r\,}$ the first term in the numerator of Eq. (58)
 dominates in reflection , such a case
 ${\cal R}\simeq 1$. If $\gamma\gg\tilde\gamma_{\,r\,}$, the first term becomes small
and the functions $v_{\,1\,}$ are essential, as well as the term
 $\sim\Omega$, which produces some asymmetry of the reflection peak.

The frequency dependence of reflection is represented in Fig. 5, one can see the
sharp asymmetry of peaks and the non-monotonic dependence ${\cal R}_{\,max\,}$ from
 $\kappa\,d$. The non-monotone is determined by the value and sign of the function
$Y_{\,1\,}$ из (62). For instance, $Y_{\,1\,}$  changes from  $Y_{\,1\,}=0.507$
 (the curve  $\kappa\,d=1.5$ in Fig.5а) up to $Y_{\,1\,}=0.072$
 (the curve $\kappa\,d=3$ ). In the last case the contribution  $Y_1\Omega$
 in the peak form is small, the asymmetry is indistinguishable,
 since the first term in Eq. (58) dominates.

 One can see the same in Fig.5b, but there $Y_1<0$ and  $\Omega>0$ corresponds to the peak maximum.
 In the case $\gamma_{\,r\,}\to 0$
 \begin{equation}
 \label{64}
 {\cal R}=\frac{v_{\,1\,}}{v}=\frac{(\zeta^{\,2\,}-1)^{\,2\,}
 \sin^{\,2\,}\kappa\,d}
 {4\zeta^{\,2\,}\cos^{\,2\,}\kappa\,d+(\zeta^{\,2\,}+1)^{\,2\,}
 \sin^{\,2\,}\kappa\,d}
 \end{equation}
and corresponds to reflection from a plane transparent layer inserted in the matter
with the different refraction index. Comparing Fig. 2 to Fig. 5 one can conclude that
the medium heterogeneity leads to the more sharp dependence of reflection from the
parameter $\kappa\,d$.

 Absorption $\cal A$ and transmission $\cal T$ are expressed as
 \begin{equation}
 \label{65}
{\cal A}=\frac{4\,\zeta\,[\,\zeta^{\,2\,}+1+
(\zeta^{\,2\,}-1)\cos\kappa\,d\,]\,
\gamma\,\tilde\gamma_{\,r\,}}{|\Delta|^{\,2\,}\,
(\Omega^{\,2\,}+\Gamma^{\,2\,}/4)},
 \end{equation}
 \begin{equation}
 \label{66}
 {\cal T}=
 \frac{4\,\zeta^{\,2\,}\,(\Omega^{\,2\,}+\gamma^{\,2\,}/4)}
 {|\Delta|^{\,2\,}\,(\Omega^{\,2\,}+\Gamma^{\,2\,}/4)},
 \end{equation}
 which transmit into Eq. (56) and Eq. (57) in the limiting cases
  $\zeta\,=\,1$ и $\kappa\,d\,=\,0$.

As it was mentioned the function $v_{\,1\,}$, which is connected with reflection of
the QW boundaries, and the term $\sim\Omega$ in the numerator are essential in forming
of the reflection curves. They determine the strong shift of the peak maximum and the
minimum appearance. On the other hand the values $\cal A$ from Eq. (65) and
 $\cal T$ from Eq. (67) coincide in their form with those  for the case of the
 homogenous medium: the difference is an appearance of the factors which do not
 depend on $\Omega$ and weakly depend on $\kappa d$. Therefore the medium
 heterogeneity influences absorption and transmission more strongly than reflection.

On the base of our analysis one can make a general conclusion that the spatial
dispersion of the electromagnetic waves and the medium heterogeneities  influence
reflection most strongly, changing radically the peak shape. These changes are most
observable in the limiting case
 $\gamma\gg\tilde\gamma_{\,r\,}$, when ${\cal R}_{\,max\,}
 \simeq (\tilde\gamma_{\,r\,}/\gamma)^{\,2\,}$.
This is since the function  $v_{\,1\,}$ from Eq. (60) and the term linear on $\Omega$
in Eq. (58) are small and influence the first term only when it is small. In another
limiting case
 $\gamma\ll\tilde\gamma_{\,r\,}\ {\cal R}_{\,max\,}\simeq 1$
and their influence is practically inobservable. If one takes into account only the
spatial dispersion or only the medium heterogeneity reflection changes comparatively
 weakly, since in these limiting cases
 $v_{\,1\,}=Y= Y_{\,1\,}=0$. The
spatial dispersion and the medium heterogeneity influence weakly the values
 $\cal A$ и $\cal T$ at $\gamma\gg\tilde\gamma_{\,r\,}$. Indeed, as it follows from
 Eq. (65) in this case
 ${\cal T}\simeq 1$  and strong change of the small value
 ${\cal R}_{\,max\,}$ influence weakly on ${\cal T}$. The same is true for
  ${\cal A}_{\,max\,}\gg{\cal R}_{\,max\,}$.

 The dependence $\cal R$, $\cal A$ и $\cal T$ on the parameter $\kappa d$,
characterizing the wave spatial dispersion in a QW, is obtained for the rectangular
QWs and infinitely high barriers. In  real semiconductor heterostructures  impurity
electrons of barriers, overflowing into the QW, distort its rectangular form near the
boundaries. Therefore our theory is applicable for clean matters and wide QW's, when
sizes of distorted regions are small in comparison to the QW width. Besides the
theory is applicable for deep QW's in which positions of energy levels and
corresponding wave functions weakly differ on energy levels and wave functions of the
infinitely deep QW. Our one-level approximation supposes, that the energy distance
between neighbor levels is mach more than the level broadening. That leads to the
restriction for the QW width. For example, for $d=500 A$ and $m_{\,c\,}=0.06m_{\,0\,}$
the energy distance of the lowest size quantized energy levels is $\simeq
10^{\,-\,3\,}eV$.

Our results are applicable in the case of a weak influence of the Coulomb interaction
on the energy spectrum of light created EHPs. These corrections are small\cite{21,25}
if

 \begin{equation}
 \label{67}
 a_{\,exc\,}^{\,2\,}\gg a_{\,H\,}^{\,2\,}, \qquad a_{\,exc\,}\gg d,
 \end{equation}
 where $a_{\,H\,}$ is the magnetic length, and the Wannier-Mott exciton radius
in absence of magnetic field is $a_{\,exc\,}=
 \hbar^{\,2\,}\epsilon_{\,0\,}/\mu\,e^{\,2\,}$.
The firs inequality of Eq. (67) is satisfied in strong enough magnetic fields, and
the second is satisfied better for QW's with large values of the dielectric function
and small reduced effective masses $\mu$ of the electron and hole. The second
condition of Eq. (67) for GaAs is satisfied at $d\leq 150 A$, when the spatial
dispersion and matter heterogeneities influence comparatively weakly on investigated
values. That is seen in Fig. 5a where the curve
 $\kappa d=0.175$ corresponds to the GaAs QW
with the width  $d\simeq 62 A$. The second inequality is approximately satisfied, but
the shift and the reflection peak asymmetry are small. If the second inequality of Eq.
 (67) is not satisfied, the dependence of the wave function on
 $z$ cannot be represented by Eq. (14). However the excitonic effect does not change
 principally our results: it influence only the radiative broadening $\gamma_r$
 of the electronic excitation. The same is true for excitonic energy levels
 in absence of magnetic field.

\section{Acknowledgements}
        S.T.P thanks the Zacatecas Autonomous University and the National
Council of Science and Technology (CONACyT) of Mexico for the financial support and
hospitality. D.A.C.S. thanks CONACyT (27736-E) for the financial support.
       This work has been partially supported by the Russian
Foundation for Basic Research and by the Program "Solid State Nanostructures Physics".



\begin{figure}
 \caption{ The functions $\varepsilon^\prime$ and $\varepsilon^{\prime\prime}$, determining
 the width changes and the peak shifts of reflection
 , transmission and absorption in the homogeneous matter. The spatial dispersion is
 taken into account. $m_{\,c\,}\,(m_{\,v\,})$ are the size quantization quantum numbers
 of electrons (holes).}
\end{figure}
\begin{figure}
 \caption{ Influence of the spatial dispersion
 on the frequency dependence of reflection
 ${\cal R}$ for the homogeneous medium. $\zeta =1,\,\gamma
/\gamma_{\,r\,}=10,\,\gamma_r=10^{-4}eV,\  m_{\,c\,}=m_{\,v\,}=1$.}
  \end{figure}

\begin{figure}
 \caption{
 Influence of the spatial dispersion
 on the frequency dependence of reflection $\cal R$
(curves 1-3) and transmission $\cal T$ (curves 4-6) for the homogeneous medium $\zeta
=1,/,\gamma/\gamma_{\,r\,}=0.1,\ \gamma_r=10^{-4}eV,\  m_c=m_v=1$. The curves $1,4 -
\kappa d=0;\  2,5 - \kappa d=1.5;\  3,6 - \kappa d=3$.}
\end{figure}
\begin{figure}
 \caption{Influence of the spatial dispersion
 on the frequency dependence of absorption $\cal
A$ for the homogeneous medium. $\zeta =1, \gamma_r=10^{-4}eV, \ m_c=m_v=1, \ a -
\gamma\gg\gamma_r,\ b - \gamma\ll\gamma_r$.}
\end{figure}
\begin{figure}
 \caption{The frequency dependence of reflection $\cal R$. The light spatial dispersion
 and the medium heterogeneities  are taken into account.
 $\gamma\gg\gamma_{\,r\,},\ \gamma_r=10^{-4},
\ m_c=m_v=1,\ a - \zeta>1,$ the curve $\kappa d=0.175$ corresponds to $GaAs, \ b -
\zeta<1.$}
\end{figure}

\end{document}